\documentclass[12pt]{article}
\usepackage{amssymb}
\usepackage{epsfig}
\usepackage[english]{babel}
\usepackage{float}
\usepackage{slashed}
\newcommand{\be}{\begin{equation}}
\newcommand{\ee}{\end{equation}}
\newcommand{\bea}{\begin{eqnarray}}
\newcommand{\eea}{\end{eqnarray}}

\def\p{\partial}
\def\pslash{\p\raise.3ex \hbox{\kern-.5em /}}
\def\delslash{\nabla\raise.3ex \hbox{\kern-.7em /}}

\begin{document}
\vskip 5cm

\begin{center}
\Large{ \textbf{\cal On limitation  of mass spectrum in
non-Hermitian $\cal PT$-symmetric models with the
$\gamma_5$-dependent mass term }}
\end{center}
\vskip 0.5cm\begin{center} \Large{V.N.Rodionov}
\end{center}
\vskip 0.5cm
\begin{center}
{Plekhanov Russian University, Moscow, Russia,  \em E-mail
vnrodionov@mtu-net.ru}
\end{center}

\begin{center}

\abstract{ The modified Dirac equations for the massive particles
with the replacement of the physical mass $m$  with the help of
the relation  $m\rightarrow m_1 + \gamma_5 m_2$  are investigated.
It is shown that for a free fermion theory with a $\gamma_5$ mass
term,  the finiteness of the mass spectrum   at the value $
m_{max}= {m_1}^2/2m_2$ takes place. In this case the region of the
unbroken $\cal PT$-symmetry may be expressed by means of the
simple restriction of the physical mass $m\leq m_{max}$.
Furthermore, we have that  the areas of unbroken $\cal
PT$-symmetry $m_1\geq m_2\geq 0$, which guarantees the reality
values of the physical mass $m$, consists of three different
parametric subregions:
 i) $0\leq m_2 < m_1/\sqrt{2}$, \,\,ii) $m_2=m_1/\sqrt{2}=m_{max},$
 \,\,(iii)$m_1/\sqrt{2}< m_2 \leq m_1$.  It is vary important, that only the first subregion (i) defined
     mass values
 $m_1,m_2,$ which correspond to the description  of  traditional particles  in  the modified
 models, because this area contain the possibility transform the modified  model to the ordinary Dirac
 theory. The second condition (ii)
is defined the "maximon" - the particle with maximal mass
$m=m_{max}$.
 In the case (iii)  we have to do with  the unusual or "exotic" particles
for description of which Hamiltonians and equations of motion have
no a Hermitian  limit.
 The formulated criterions  may be
 used as  a major test in the process of the division of
 considered models into  ordinary and "exotic fermion theories". }

\end{center}


\section{Introductory remarks}

As it is well-known the idea about existence of a maximal mass in
a broad spectrum mass of elementary particles at the  Planck mass
was suggested by Moisey Markov in 1965 \cite{Markov}
$$
m\leq m_{Planck}\cong 10^{19} GeV.
$$
  After  that  a more radical approach was developed by V.G.
Kadyshevsky \cite{Kad1}.   His model  contained a limiting mass
$M$ as a new fundamental physical constant. Doing this condition
of finiteness  of the mass spectrum should be introduced by the
relation: \be\label{Mfund1} m \leq M,\ee where a new constant $M$
was named by the \emph{fundamental mass}.

 Really in the papers
\cite{Kad1}-\cite{Rod} the existence of mass $M$ has been
understood as a new principle of Nature similar to the
relativistic and quantum postulates, which was put into the ground
of the new quantum field  theory. At the same time the new
constant $M$ is introduced in a purely geometric way, like the
velocity of light is the maximal velocity in the special
relativity.

Indeed, if one chooses a geometrical formulation of the quantum
field theory, the adequate realization of the limiting mass
hypothesis  is reduced to the choice of the de Sitter geometry as
the geometry of the 4-momentum  space of the constant curvature
with a radius equal to $M$ \cite{Kad1}. The detailed analysis of
the different aspects of the construction of the modified quantum
field theory with the maximal mass in the curved momentum de
Sitter space has allowed to obtain a number of interesting
results. In particular, it has been shown that  non-Hermitian
fermionic Hamiltonians with the $\gamma_5$-dependent mass term
must arise in the modified theory (see, for
example,\cite{Max},\cite{KMRS} ).

Now it is well-known fact the reality of the spectrum in models
with a non-Hermitian Hamiltonian is a consequence of $\cal PT$
-invariance of the theory, i.e. a combination of spatial and
temporary parity of the total Hamiltonian: $[H,{\cal PT}]\psi =0$.
When the $\cal PT$ symmetry is unbroken, the spectrum of the
quantum theory is real. These results explain the growing interest
in this problem. For the past a few years studied a lot of new
non-Hermitian $\cal PT$-invariant systems (see, for example,
\cite{Rod1} - \cite{neznamov2}).  In the literature, which was
devoted to the study of non-Hermitian operators there are
examples, with the $\gamma_5$ mass extension.

In particular the modified Dirac equations for the massive
Thirring model in two-dimensional space-time with the purely
algebraic replacement of the physical mass $m$ by $m\rightarrow
m_1+ \gamma_5 m_2$ (where $m_1\geq 0$, and $m_2$ real) was
investigated by Bender et al.\cite{ft12}. As the foundation of
this study  is assumed the a model with the density of the
Hamiltonian, which is represented in the form:

\be\label{D1} {\cal
H}(x,t)=\bar{\psi}(x,t)\Big(-i\overrightarrow{\partial}\overrightarrow{\gamma}+m_1+\gamma_5
m_2 \Big)\psi(x,t). \ee The equation of motion  following from the
(\ref{D1}), may be expressed  as

\be\label{D2} \Big(i\partial_\mu\gamma^{\mu}-m_1-\gamma_5 m_2
\Big) \psi(x,t)=0. \ee It is obvious that the Hamiltonian
associated with the Hamiltonian density (\ref{D1}) is
non-Hermitian due to the appearance in it the $\gamma_5$-dependent
mass components ($H\neq H^{+}$).

A.Mustafazadeh identified the necessary and sufficient
requirements of reality  of eigenvalues for pseudo-Hermitian and
$\cal PT$-symmetric Hamiltonians and formalized the use these
Hamiltonians in his papers \cite{alir},\cite{ali}. According to
the  recommendations of this works we can define  Hermitian
operator $\eta$, which transform non-Hermitian  Hamiltonian
 by means of invertible
transformation to the Hermitian-conjugated one. It is easy to see
that with Hermitian operator \be \label{eta} \eta = e^{\gamma_5
\vartheta} \ee we can obtain
 \be\label{D3}  \eta H \eta^{-1}= H^{+},\ee
  \be\label{H} H =\alpha p+ \beta(m_1
+\gamma_5 m_2)\ee  and \be\label{H+} H^+ =\alpha p+ \beta(m_1
-\gamma_5 m_2),\ee where  matrices
$\alpha_i=\gamma_0\cdot\gamma_i$ and $\beta=\gamma_0$, and
$\vartheta= \textmd{arctanh} (m_2/m_1)$

In addition, multiplying the Hamilton operator $H$ on the left on
$ e^{{\vartheta\gamma_5 }/2}$, we can obtain

\be \label{H0}  e^{{\gamma_5 \vartheta}/2} H = H_0  e^{{\gamma_5
\vartheta}/2},\ee where $H_0 =\alpha p +\beta m $ is a ordinary
Hermitian Hamiltonian of a free particle.

The mathematical sense of the action of the operator (\ref{eta})
it turns out, if we notice that according to  the properties of
matrices $\gamma_5$, all the even  degree of $\gamma_5$ are equal
to 1, and all odd degree are equal to $\gamma_5$. Given that $ch$
decomposes on even and $sh$ odd degrees, the expressions
(\ref{D3})-(\ref{H0}) can be obtained by representing exponential
operator $\eta$ in the form
$$
\eta=e^{\gamma_5 \vartheta}=\textmd{ch}\vartheta+\gamma_5
\textmd{sh}\vartheta,
$$
where \be\label{chsh} \textmd{ch}\vartheta =
m_1/m;\,\,\,\,\textmd{sh}\vartheta = m_2/m, \ee
 and taking into
account that the matrices $\gamma_5$ commute with matrices
$\alpha_i$ and anti-commute with $\beta$.

 The region of the unbroken
$\cal PT$-symmetry in the paper \cite{ft12} has been found  in the
form

\be\label{Thir} {m_1}^2\geq {m_2}^2. \ee

However, it is not apparent that the area with undisturbed $\cal
PT$-symmetry (\ref{Thir}) does not  include  the regions,
corresponding to the  some unusual particles, description of which
radically distinguish from traditional one. This particles which
was named  "exotic" in the frame of geometric approach to the
construction of model with the Maximal Mass, should play a special
role in the
  world of the elementary particles  \cite{Max},\cite{KMRS}.

  The distinguishing feature of
"exotic"  particles consists of the fact that Hamiltonians and
equations of motion  for their description have no the limit when
$ M\rightarrow \infty $. Thus,  they  not to agree in this limit
with  the ordinary Dirac expressions and one can assume that in
this case we deal with a description of some new particles,
properties of which have not yet been studied. This fact for the
first time  has been fixed back in the early works on the
development of the theory with a fundamental mass \cite{Kad1}.

In the frame of   purely algebraic approach, when take place the
extending of mass parameter  $m\rightarrow m_1+\gamma_5 m_2$,
there are not  an explicit  restrictions  of masses. However, from
(5),(6) one can  easy obtain a number of the limiting values of
the mass parameters. In this connection it is very
  interesting to learn could  in algebraic model receive a
  description of such particles? Consequently the question arises:
  "what precisely particles are considered: exotic or traditional, when the
condition (\ref{Thir}) is executed?"

 It is important  to note that the previous works in
pseudo-Hermitian quantum mechanics \cite{ft12},\cite{ft13} with
$\gamma_5$-dependent mass term have been carried out in such a
way, that condition (\ref{Thir}) is \textbf{a single inequality}
for parameters of mass in the model. The main result obtained by
us consists of that  the region (\ref{Thir}) of unbroken $\cal PT$
symmetry in reality has an internal structure where implicated as
ordinary and "exotic" particles. This conclusion managed to get
thanks to the existence of the restriction of the mass spectrum of
particles $m \leq m_{max}$ automatically arising in relativistic
model with $\gamma_5$-dependent mass term in the Hamiltonians
(\ref{H}) and (\ref{H+}).

This paper has the following structure. In section II the
\textbf{necessary} and \textbf{sufficient conditions} of
restrictions the mass spectrum are formulated  in considered
model. In the third section we study the basic characteristics of
$\cal PT$-symmetric free fermionic models with $\gamma_5$  massive
term.

\section{Necessary and sufficient conditions  of the mass
spectrum restrictions  in the model with $\gamma_5$ mass term. }

Consider now the algebraic approach developed in the numerous
papers on the study of quantum non-Hermitian mechanics.
 It was possible to expect, that the appearance of the models
described by Hamiltonians  type (\ref{H}),(\ref{H+}) is the
prerogative of a purely geometric approach to the construction of
a modified theory with a maximal mass. However, in the paper
 \cite{ft12} was considered the ${\cal PT}$-symmetric
{massive Thirring model } with $\gamma_5$ mass extension in
(1+1)-dimensional space.

Let us consider  the relativistic quantum mechanics with the
$\gamma_5$-mass term in the case of 3+1 dimensional space-time. We
introduce the following representation of a $\gamma$-matrices
\cite{2D}: \be \gamma_0=\left(\begin{array}{cc}0 & I\\ I & 0
\end{array}\right) \quad \quad
\vec{\gamma}=\left(\begin{array}{cc}0 & \vec{\sigma}\\
-\vec{\sigma}& 0
\end{array}\right). \label{e14} \ee According to these
definitions, $\gamma_0^2=1$ and $ {\gamma_i}^2=-1$. We also have
\be \gamma_5= =\left(\begin{array}{cc}I&0\\0&-I\end{array}\right),
\label{e15} \ee so that $\gamma_5^2=1$.

In this regard, there is interest in the study of the parameters,
characterizing of the masses, included in the $\gamma_5$-dependent
mass model. Note that the $\gamma_5$ - extension of the mass in
the Dirac equation consists of replacing $m\rightarrow
m_1+\gamma_5 m_2$ and after that two new mass parameters: $m_1$
and $m_2$ arises.

First-order equations (\ref{D2}) can be transformed into equations
of second order by applying to (\ref{D2}) operator: \be\label{D22}
\Big(i\partial_\mu\gamma^{\mu}+ m_1-\gamma_5 m_2 \Big).\ee In a
result the Dirac equation converts  to the Klein-Gordon equation:
\be \label{KG} \left(\partial^2+m^2\right)\psi(x,t)=0 \label{e20}
\ee where \be \label{012} m^2={m_1}^2- {m_2}^2. \ee It is easy to
see from (\ref{012}) that the physical mass $m$, appearing in the
equation (\ref{KG}), is real when the inequality \be \label{e210}
{m_1}^2\geq {m_2}^2.\ee is accomplished.

Developed  algebraic formalism in \cite{ft12}, no contains
indications of the existence of other restrictions, in which
participates the mass, in addition to (\ref{e210}). However, it is
 important, that this conditions should be obtained because
they  can help to establish the connection between algebraic and
geometric approaches. Note also that  the existence of such
restrictions, in particular, may essentially modify  the
fundamental results obtained in the paper \cite{ft12}.

 Consider the Hamiltonians of the type (\ref{H}) and
show that the the algebraic approach allows one to set the
\textbf{sufficient condition} of the limitations of the mass
spectrum of particles in relativistic quantum mechanics with
$\gamma_5$-mass component. As noted above, the inequality
(\ref{e210}) was considered  \cite{ft12} as  the
 requirement that determines the presence of broken or
unbroken  $\cal PT$ - symmetry of the Hamiltonian. However, it is
easy to see that (\ref{e210}) may not be as the single condition
of this type.

Writing the following obvious inequality: \be (m-m_2)^2 \geq 0 \ee
and taking into account (\ref{012}), we can obtain \be \label{M}m
\leq \frac{{m_1}^2}{2 m_2}= m_{max},\ee that is direct indication
of the existence of the  restriction of the physical mass $m$ in
the considered model with ${\gamma_5}$-dependent mass term.

It is interesting that (\ref{M}) is obtained as the result of the
simple algebraic transformation of relationships with subsidiary
parameters of mass $m_1$ and $m_2$. It is quite natural that the
value of the $m_{max}$ is expressed through a combination of them.
In particular, as the degree of deviation of the Hamiltonian $H$
of a Hermitian forms is characterized by the mass  $m_2$, then its
value can be expressed from (\ref{M}) \be\label{m2} m_2=
\frac{{m_1}^2}{2 m_{max}}.\ee To see whether the modified
Hamiltonian (\ref{H}) is Hermitian, we must check  whether the
contribution of the mass $m_2$ becomes vanishingly small? We do so
by  doing the parameter maximal mass very large (formally, when
$m_{max}\rightarrow \infty$).

 Following the  (\ref{chsh}) and taking advantage of the
symbol \be \label {m21}m_2=m_1 \textmd{tanh}\vartheta,\ee we can
obtain \be\label{m211} \frac{m_1}{2
m_{max}}=\textmd{tanh}\vartheta\leq 1,\ee
 Then we can  establish the limits of change of
parameters.  At preset values of $m$ and $m_{max}$ as it follows
from the (\ref{M})-(\ref{m211})  the limits of variation of
parameters $m_1$ and $m_2$ are the following: \be\label{limit}
m\leq m_1\leq 2 m_{max}; -2m_{max} \leq m_2 \leq 2m_{max},\ee
where considering the possible change of the sign of $m_2$. In the
areas of change of these parameters has a point in which
$\textmd{tanh}\vartheta_0=1/\sqrt{2}$, where we have
\be\label{m111} m_1=\sqrt{2}m_{max}; m_2=m_{max}.\ee In this point
the physical mass $m$ (see (\ref{M})) reaches its maximum value
$m=m_{max}$.

   Thus,
we may rewrite (\ref{H}) in the resulting form \be\label{eps} H=
\vec{\hat{\alpha}}\vec{p}+\hat{\beta} m_1\Big(1+ \gamma_5
\textmd{tanh}\vartheta\Big),\ee and taking into account
(\ref{m111})
 the areas of  possible values of the parameter $\vartheta$ can
 be presented in the form:
  (i) $0\leq\vartheta < \vartheta_0$ -
corresponds to the theory  having a Hermitian limit, i.e. the
limit, when $m_2 \rightarrow 0$ ($ m_{max}$ tends to infinity)
exists,\footnote{In the case of geometric approach this limit was
called a "flat limit" because when the parameter $M$ is very large
the anti de Sitter geometry does not differ from the Minkowski
geometry in four dimensional pseudo-Euclidean p-space} and (ii)
$\vartheta_0<\vartheta <\infty$ - refers to the occasion when the
such a limit is absent. The first condition corresponds to the
description of ordinary particles and the second - to unusual or
exotic particles. The limit value $\vartheta = \vartheta_0 =
arctanh(1/\sqrt{2}) $ is responsible for the particles with the
maximal mass, which was named by the maximons \cite{Markov}.

Thus, the limitation of the mass spectrum of particles (\ref{M}),
described by Hamiltonians (\ref{H}),(\ref{H+}) and (\ref{eps}) is
a simple consequence of the $\gamma_5$-mass extension in
(\ref{D1}),(\ref{D2}). Therefor the presence of the non-Hermitian
Hamiltonian with a $\gamma_5$-mass contribution,  in essence, can
be interpret as the \textbf{sufficient condition} of the
limitation of the mass spectrum of particles in fermion models
with $\gamma_5$ - massive term.

On the other hand, in the framework of geometrical approach, when
the basis of the modified QFT with the Maximal Mass is a postulate
(\ref{Mfund1}) we need to get  the appearance of non-Hermitian
Hamiltonians with a $\gamma_5$-contribution in its fermion sector
(see for example \cite{Max},\cite{KMRS},\cite{Rod1}). This
consequence can be interpreted \textbf{as a necessary condition }
of the finiteness of the mass spectrum \cite{Markov}.

Thus, we can see that the \textbf{necessary and a sufficient
conditions} of the limitation of the physical mass of  particles
(\ref{Mfund1}) and (\ref{M})  in fermion sector are the using of
the non-Hermitian
 $ \cal PT$ - symmetric quantum models with a $\gamma_5$
- massive contribution.

\section{Free fermion models with $\gamma_5$-massive
contributions and  the areas of unbroken $\cal PT$ symmetry}

 Condition
(\ref{e210})  is executed automatically, if one enter the
parametrization (\ref{chsh}). Moreover, from (\ref{M})
 we can express the values of
$m,\,\,m_1,\,\,m_2$ with parameter $\vartheta$.

\begin{figure}[h]
\vspace{-0.2cm} \centering
\includegraphics[angle=0, scale=0.5]{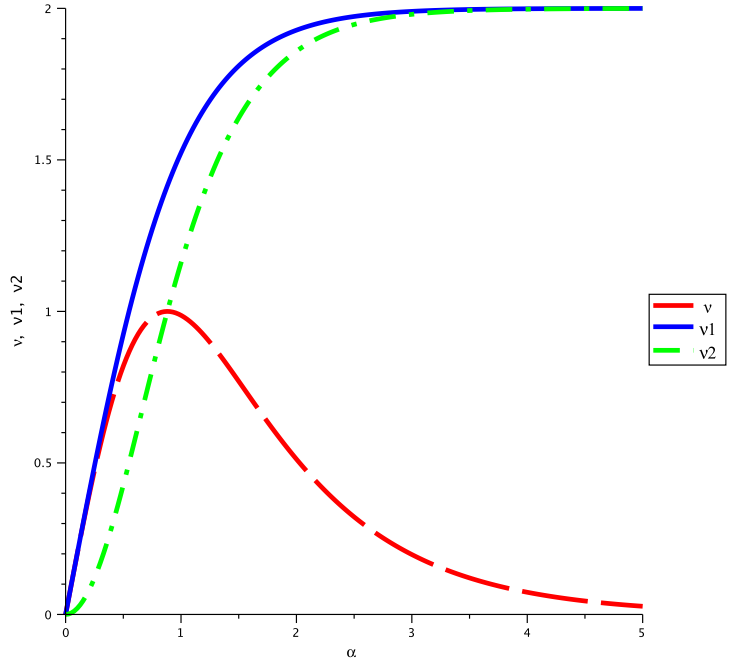}
\caption{Dependence of $\nu=m/m_{max}, \nu_1=m_1/m_{max},
\nu_2=m_2/m_{max} $ from the parameter $\vartheta $}
\vspace{-0.1cm}\label{Fig.1}
\end{figure}

 Fig. 1 shows the dependence of the relative values
$\nu=m/m_{max}$, $\nu_1=m_1/m_{max}$ and $\nu_2=m_2/m_{max}$ on
the parameter  $\vartheta$. The values of the parameters
$\nu1,\,\nu2$ are the following: $\nu\leq\nu_1\leq 2$,
$0\leq\nu_2\leq 2$.  A particle mass $\nu$ may vary in a range of
$0\leq m\leq m_{max}$. When $\vartheta_0 = 0.881 $ it reaches its
maximum, which corresponds to the maximon $m=m_{max} $.

Using (\ref{M}),(\ref{chsh})  we can also obtain  \be\label{tan1}
\tanh(\vartheta) = \sqrt{\frac{1 \pm\sqrt{1-\nu^2}}{2}}. \ee Two
signs of the root in (\ref {tan1}) are interpreted as  a
two-valued variables   $\nu_1(\nu_3)$ and $\nu_2(\nu_4)$, which
are multi-valued functions of $\nu$. Thus, we have

\be\label{m1} \nu_1(\nu_3) = \sqrt{2} \sqrt{1 \mp\sqrt{1-\nu^2}};
 \ee

\be \label{m2}\nu_2(\nu_4) = \left( 1 \mp\sqrt{1-\nu^2}\right).
\ee

It is easy to see, that between the  symbols for the masses
 in geometric approach \cite{Kad1} and
obtained here the values of (\ref{m1}), (\ref{m2}), after
identification of the limiting masses $M$ and $m_{max}$, there are
simple correlations.



Fig. 2 demonstrates the dependence of the parameters
$\nu_1(\nu_3)$, $\nu_2(\nu_4)$ on the variable $\nu$. Thus, the
region of the existence of unbroken  $\cal PT$ symmetry can be
represented in the form $0 \leq\nu\leq 1$. For these values of
$\nu$ parameters $\nu_1$ and $\nu_2$ determine the masses of the
modified Dirac equation with a maximal mass $m_{max}$, describing
the  particles having the actual mass $m\leq m_{max}$. However,
the new Dirac equations nonequivalent, because one of them
describes ordinary particles ($\nu_1,\nu_2$), and the other
corresponds to their exotic partners($\nu_3,\nu_4$). The special
case of Hermiticity is on the line $\nu=1$ ($m=m_{max}$ - the case
of the maximon), which is the boundary of the unbroken $\cal PT$ -
symmetry. In this point of the plot we have $\nu_1=\nu_3=\sqrt{2}$
and $\nu_2=\nu_4=1$.

\begin{figure}[h]
\vspace{-0.2cm} \centering
\includegraphics[angle=0, scale=0.5]{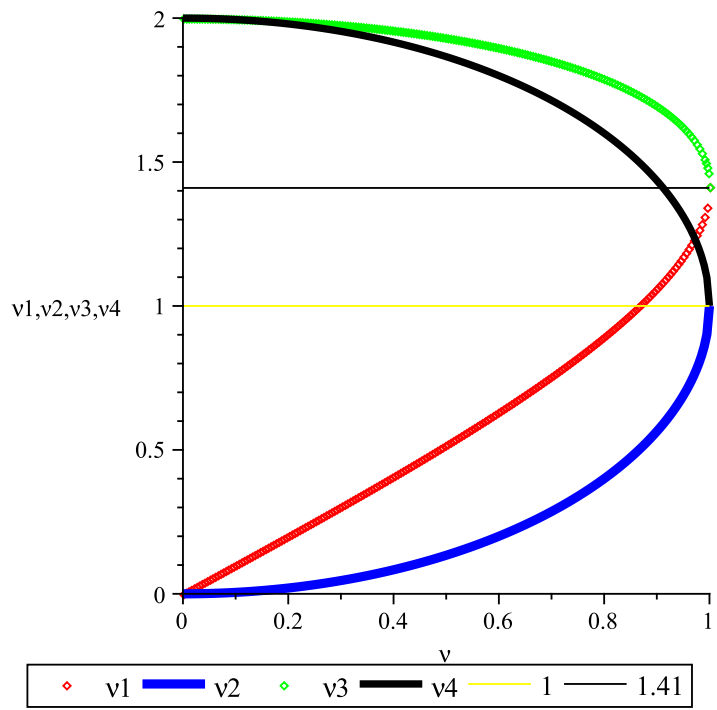}
\caption{The values of parameters   $\nu_1, \nu_2,  \nu_3, \nu_4$
as the function of $\nu$} \vspace{-0.1cm}\label{f4}
\end{figure}

\begin{figure}[h]
\vspace{-0.2cm} \smallskip
\includegraphics[angle=0, scale=0.5]{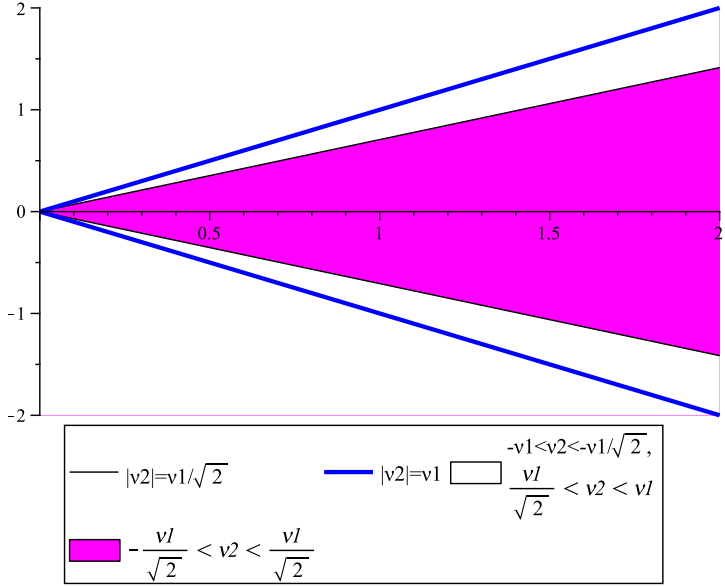}
\caption{The parametric areas of the unbroken $\cal PT$-symmetry
${\nu_1}^2 \geq {\nu_2}^2$ in plane $\nu_1$,$\nu_2$ for the
Hamiltonian (\ref{eps})
consists of three specific subregions. Only the shaded area $II.$
meets the ordinary particles, and the bordering with it regions
$I.$ and $III.$ correspond to the description of the exotic
fermions.} \vspace{-0.1cm}\label{f4}
\end{figure}

It is easy to verify that the new values of the mass parameters $
m_3$, $ m_4$ still satisfy the conditions (\ref{012}) and
(\ref{e210}). We emphasize once again that, from the formulas
(\ref{m1}),(\ref{m2}) in the case of the upper sign it should be
$m_1\rightarrow m$ and $m_2\rightarrow 0$ when $m_{max}
\rightarrow \infty,$ (i.e. there is a so-called Hermitian or "flat
limit", determined in the frame of the geometric approach
\cite{Max},\cite{KMRS}). However, when one choose a lower sign
(i.e.  for the $m_3$ and $m_4$) such a limit is absent.

 In the frame of the condition (\ref{Thir})\footnote
 {Note that this inequality was considered in the paper
\cite{ft12} as a expression, which is completely defines all
possible consequences of unbroken ${\cal PT}$-symmetry in this
model.} at Fig.3 we can see three specific sectors of unbroken
$\cal PT$-symmetry of the Hamiltonian (\ref{eps}) in the plane
$\nu_1, \nu_2$. The plane $\nu_1,\nu_2$ (there is considered the
possible change of the sign of the parameter $m_2$) may be divided
by the three groups of the inequalities:

$$I.\,\,\,\,\,\,\,\,\,\,\,\,\,\,\,\,\,\,\, \nu_1/\sqrt{2} \leq \nu_2 \leq\nu_1,$$
$$II.\,\,\,\,-\nu_1/\sqrt{2}< \nu_2 < \nu_1/\sqrt{2},$$
$$III. \,\,\,\,\,\,\,\, -\nu_1 \leq \nu_2\leq-\nu_1/\sqrt{2},$$

Only the area $II.$ corresponds to the description of ordinary
particles, then $I.$ and $III.$ agree with the description of some
as yet unknown particles. This conclusion is not trivial, because
in contrast to the geometric approach, where the emergence of new
unusual properties of particles associated with the presence in
the theory a new degree of freedom (sign of the fifth component of
the momentum $\varepsilon=p_5/|p_5|$ \cite{KMRS}), in the case of
a simple extension of the free Dirac equation  due to the
additional $\gamma_5$-mass term, the satisfactory explanation is
not there yet.

 Taking into account properties of matrix $\gamma_o,
\vec{\gamma}, \gamma_5$, we can  write also complex-conjugate
equation from equation (\ref{D2}) \be\label{conj}
\left(-p_0\tilde{\gamma_0}
-\textbf{p}\widetilde{\vec{\gamma}}-m_1-\gamma_5
m_2\right)\psi^{*}=0, \ee where $\tilde{\gamma_\mu }$ are
transpose matrix   and here take place the replacement of \be
\gamma p = p_0\gamma_0 - \textbf{p}\vec{\gamma} = i\gamma_0
\frac{\partial}{\partial t}+i\vec{\gamma}{\bf\nabla ,}\ee
$\vec{\gamma}= (\gamma_1,\gamma_2,\gamma_3)$.

 Rearranging function $\psi^{*}$
according to $\tilde{{\gamma^\mu}}\psi^{*}=\psi^{*}\gamma^\mu$
then multiply the equation (\ref{conj}) on the right of
$\gamma_0$. Noticing that \be \vec{\gamma}\gamma_0 = -
\gamma_0\vec{\gamma}, \,\,\,\gamma_5\gamma_0=-\gamma_0\gamma_5\ee
and introducing new bespinor $\bar{\psi}= \psi^{*}\gamma_0$, we
can obtain \be\label{bar} \bar{\psi}\left(\gamma p+m_1 -\gamma_5
m_2\right)=0.\ee The operator $p$ is assumed here acts on the
 function, standing on the left of it.
Using (\ref{chsh}) we can write equation (\ref{D2}), (\ref{bar})
in the following form \be \label{1} \left(p\gamma - m
\eta\right)\psi=0\ee \be\label{2} \bar{\psi} \left (p\gamma + m
\eta^{-1}
 \right)=0\ee

Repeating the generally accepted procedure of obtaining the
continuity equation for a Hermitian case (see,for example
\cite{LL}) we get similar results for a $\gamma_5$-modified
non-Hermitian quantum mechanics.
 Multiply (\ref{1}) on the left of the $\bar{\psi}
e^{-\vartheta\gamma_5} $ and the equation (\ref{2}) on the right
of the $e^{\vartheta\gamma_5}\psi$ and sum up the resulting
expressions, one can obtain \be
\bar{\psi}e^{-\vartheta\gamma_5/2}\gamma_{\mu}
e^{\vartheta\gamma_5/2}(p \psi) +
(p\bar{\psi})e^{-\vartheta\gamma_5
/2}\gamma_{\mu}e^{\vartheta\gamma_5/2}=p_\mu\left(\bar{\psi}e^{-\vartheta\gamma_5
/2}\gamma_{\mu}e^{\vartheta\gamma_5/2}\psi\right)=0 \ee Here
brackets indicate  which of the  function are  subjected to  the
action of  the operator $p$. The obtained equation has the
 form  of the continuity equation $\partial_\mu j_\mu =0. $ Thus, the value of \be j_\mu = \bar{\psi}e^{-\vartheta\gamma_5
/2}\gamma_\mu
 e^{\vartheta\gamma_5}\gamma_\mu\psi = \left(\psi^{*} e^{\vartheta\gamma_5}\psi, \psi^{*}\gamma_0\vec{\gamma }
 e^{\vartheta\gamma_5}\psi \right)\ee

  Thus here the value of $j_\mu $ is a 4-vector of current density of
  particles in the model with $\gamma_5$-mass extension.
   It is very important that its temporal component \be\label{j_0} j_0=\psi^{*}e^{\vartheta\gamma_5}\psi \ee
   positively defined  and does not change in time.
   It is easy to see from the following procedure.

Let us construct  the  norm of any state for considered model for
arbitrary vector, taking into account the weight operator $\eta$
(\ref{eta}):
$$
\Psi= \left( \begin{array}{cc}
 x+i y&{} \\
u+iv&{}\\
z+iw&{}\\
t+ip&{}
\end{array}\right).
$$
Using (\ref{chsh}),(\ref{e14}),(\ref{e15})   and (\ref{j_0}),  in
a result we have

$$
 \Psi^{*}\eta=\left(
\frac{m_1+m_2}{m}(x-iy),\frac{m_1+m_2}{m}(u-iv),
\frac{m_1-m_2}{m}(z-iw), \frac{m_1-m_2}{m}(t-ip)\right).
$$
Then \be \label{Psi} \langle{}
\Psi^{*}\eta|\Psi\rangle=\frac{m_1+m_2}{m}(x^2+y^2)+\frac{m_1+m_2}{m}(u^2+v^2),\frac{m_1-m_2}{m}(z^2+w^2),\frac{m_1-m_2}{m}(t^2+p^2)
\ee is explicitly non negative, because  $m_1\geq m_2$ in the area
of unbroken ${\cal PT}$-symmetry (\ref{Thir}). Given that \be \int
d^3 x\psi^{*}\eta \psi = \int d^3 x j_0 \ee we have  the
conservation of probability directly connected with correlation
(\ref{j_0}) and weight operator $\eta$.

\section{Conclusion}

Starting with the researches, presented in the previous sections,
we have shown that Dirac Hamiltonian of a particle with $\gamma_5$
- dependent mass term is non-Hermitian, and has the unbroken $\cal
PT$ - symmetry in the area ${m_1}^2\geq {m_2}^2,$ which has three
of a subregion. Indeed with the help of the algebraic
transformations we obtain a number of the consequence of the
relation (\ref{012}). In particular there is the restriction of
the particle mass in this model: $m\leq m_{max}$, were
$m_{max}={m_1}^2/2 m_2$. Outside of this area the $\cal PT$ -
symmetry of the modified Dirac Hamiltonians is broken.

 In addition, we have shown
that the introduction of the postulate about the limitations of
the mass spectrum, lying in the ground of the a geometric approach
to the development of the modified QFT (see, for example
\cite{Max},\cite{KMRS}), leads to the appearance of non-Hermitian
$\cal PT $-symmetric Hamiltonians in the fermion sector of the
model with the Maximal Mass. Thus, it is shown that  using of
non-Hermitian $\cal PT$-symmetric quantum theory with $\gamma_5$
mass term may be considered as \textbf{ necessary and sufficient
conditions} of the appearance of the limitation of the mass
particle (\ref{M}) in a fermion sector of the model.

In particular, this applies to the modified Dirac equation in
which produced the substitution $m\rightarrow m_1+ \gamma_5 m_2 $.
Into force of the ambiguity of the definition of parameters $m_1,
m_2$ the inequality $m_1\geq m_2\geq 0$ describes a particle of
two types. In the first case, it is about ordinary particles,
when $m_1,m_2\geq 0$ mass parameters  are limited by the terms

\be \label{01}     0\leq m_2 \leq m_1/\sqrt{2} . \ee

In the second area we are dealing with so-called «exotic partners»
of ordinary particles, for which is still accomplished
(\ref{e210}), but one can write

\be \label{02} m_1/\sqrt{2} \leq m_2 \leq m_1. \ee

Intriguing difference between particles of the second type from
traditional fermions is that they are described by the other
modified Dirac equations. So, if in the first case(\ref{01}), the
equation of motion there has a limit transition when $m_{max}
\rightarrow \infty$ that leads to the standard Dirac equation,
however in the inequality (\ref{02}) such a limit is not there.

Thus, it is proved that the main progress, obtained by us the in
the algebraic way of the construction of the fermion model with
$\gamma_5$ - dependent mass term applies to   the limitations of
the mass spectrum. Furthermore, the possibility of describing of
the exotic particles are turned out essentially the same as in the
model with a maximal mass, which was investigated by
V.G.Kadyshevsky with colleagues \cite{Kad1} - \cite{Rod} on the
basis of geometrical approach. It is also shown that the
transition point at the scale of
 the mass from the ordinary particles to the exotic one this is  mass of
 the maximon.

On the basis of (\ref{M}) it also has been shown that the
parameters $ m_1$ and $ m_2$ have the auxiliary nature. This fact
is easily proved by means of the comparison of  the ordinary and
exotic fermion fields. Thus, it is the important conclusion   that
the description of exotic fields may be considered  with the help
of the algebraic approach and  is not the prerogative of the
geometric formalism. Note that the polarization properties of the
exotic fermion fields fundamentally differ from  the standard
fields that with taking into consideration of interactions  may be
of interest in the future  researches. For example, if for
massless fermions in the case of ordinary particles we have the
mass condition  $m_1=m_2=0$ and  as a consequence of the Weyl
equations \be (E+\textbf{p}\vec{\sigma})u_L=0,\ee \be
(E-\textbf{p}\vec{\sigma})u_R=0, \ee  where $E=\pm p$,
 in the case of exotic massless particles we have the
possibility $m=0$, when $m_2 =m_1 =2 m_{max}$ (see Fig.2). This
leads to the modified system of  Weyl equations: \be
(E+\textbf{p}\vec{\sigma})u_L=0\ee

\be(E-\textbf{p}\vec{\sigma})u_R -4m_{max}u_L=0,\ee which remains
consistent when $E=\pm p$. It is easy to see that in this case the
approximation $m_2\rightarrow 0$ (that is equivalent to
$m_1\rightarrow 0$ and hence $ m_{max}={m_1}^2/m_2 \rightarrow 0$)
lead to the ordinary Weil equations.

The presence of this possibilities lets hope for that in Nature
indeed there are the Fundamental Mass Value and some ”exotic
fermion fields”. It is tempting to think that the quanta of the
exotic fermion field have a relation to the structure of the ”dark
matter”.

{\bf Acknowledgment:} We are grateful to Prof. V.G.Kadyshevsky for
fruitful and highly useful discussions.

 \end{document}